\documentclass[a4paper]{article}

\usepackage{enumitem}% used
\usepackage{float}
\usepackage{amsfonts}
\usepackage{algorithm}
\usepackage{xcolor}
\usepackage{titling} % spacing above title
\usepackage{multirow}

\usepackage[numbib]{tocbibind} % to add number to references section

\usepackage{amsmath} %for dealing with mathematics,
\usepackage{amsthm} %for dealing with theorem environments,
\usepackage{cite} %for dealing with citations
\usepackage{hyperref} %for linking the cross references
\usepackage{graphicx}
\usepackage{algorithmic} %for describing algorithms
\usepackage{authblk}
\usepackage{multicol}
\usepackage{blindtext}
\usepackage{epstopdf}
\usepackage[margin=1.2cm, bottom=2cm]{geometry}

\theoremstyle{plain}
\theoremstyle{definition}
\newtheorem{theorem}{Theorem}
\newtheorem{definition}{Definition}

\begin{document}

% \title{Privacy-Preserving Quantum Neural Networks for Probabilistic Optimal Power Flow Approximation}
\title{A Differentially Private Quantum Neural Network for Probabilistic Optimal Power Flow}
% \vspace{-5ex}
\author[1]{Yuji Cao}
\author[1*]{Yue Chen}
\author[2]{Yan Xu}

\affil[1]{Department of Mechanical and Automation Engineering, The Chinese University of Hong Kong, Hong Kong, China.}
\affil[2]{School of Electrical and Electronic Engineering, Nanyang Technological University, Singapore 639798, Singapore.\newline *Email: yuechen@mae.cuhk.edu.hk}

\date{\vspace{-5ex}}

\setcounter{Maxaffil}{0}
\renewcommand\Affilfont{\itshape\small}
\maketitle

\begin{changemargin}{1.2cm}{1.2cm} 
Keywords: Probabilistic Optimal Power Flow; Quantum Neural Network; Variational Quantum Circuit; Differential Privacy.
\end{changemargin}

\begin{abstract}
The stochastic nature of renewable energy and load demand requires efficient and accurate solutions for probabilistic optimal power flow (OPF). Quantum neural networks (QNNs), which combine quantum computing and machine learning, offer computational advantages in approximating OPF by effectively handling high-dimensional data. However, adversaries with access to non-private OPF solutions can potentially infer sensitive load demand patterns, raising significant privacy concerns. To address this issue, we propose a privacy-preserving QNN model for probabilistic OPF approximation. By incorporating Gaussian noise into the training process, the learning algorithm achieves ($\varepsilon, \delta$)-differential privacy with theoretical guarantees. Moreover, we develop a strongly entangled quantum state to enhance the nonlinearity expressiveness of the QNN. Experimental results demonstrate that the proposed method successfully prevents privacy leakage without compromising the statistical properties of probabilistic OPF. Moreover, compared to classical private neural networks, the QNN reduces the number of parameters by 90\% while achieving significantly higher accuracy and greater stability.
\end{abstract}

\begin{multicols*}{2}

\raggedcolumns % not balance two columns

\section{Introduction}
\label{sec:intro}

The integration of renewable energy sources into power systems introduces significant variability and uncertainty in power generation and demand. Probabilistic optimal power flow (OPF) is essential for managing these uncertainties by determining optimal operating points that account for the stochastic nature of renewables and loads~\cite{li2008analysis}. However, solving probabilistic OPF problems is computationally intensive, calling for methods to provide fast and accurate solutions for real-time applications.

% quantum computing computational advantages -> QML
Learning-based methods have emerged as effective tools for accurately approximating OPF solutions due to their ability to capture complex nonlinear relationships from the data~\cite{huang2021deepopf}. As power system complexities continue to grow, there is a pressing need for more advanced computational approaches. Building on the exponential speedup provided by quantum computing, the rapid advancements in quantum machine learning (QML) offer promising prospects for developing scalable and efficient power system computations~\cite{ajagekar2020quantum,liu2024towards}. Indeed, quantum computing uses superposition and entanglement to achieve efficient data representation and parallel computation. Hence, it holds great potential to facilitate large-scale computation in next-generation power systems. Moreover, the inherent speed advantages of quantum computing allow QML methods to perform computations more quickly, thereby supporting real-time optimization and decision-making.

%TODO： 从电网角度举例，可以通过电压判断用户行为等... 
Despite their computational efficiency, learning directly from the non-private OPF solutions can lead to privacy breaches, as the non-private voltage and power flow information may reveal sensitive load patterns~\cite{wang2020privacy}. By introducing calibrated noise into the learning process, differential privacy has recently become a promising tool for privacy protection~\cite{dwork2006differential}. Mathematically, differential privacy provides a theoretical framework to quantify and control privacy losses, based on which one can design an expected trade-off between model utility and privacy preservation. 

Differentially private OPF problems have been extensively investigated. To the best of our knowledge, most existing works are based on optimization methods~\cite{dvorkin2020differentially, lei2024decentralized}, while learning-based methods are seldom explored. These two approaches differ greatly in the noise injection mechanisms. In learning-based methods, noise is incorporated in gradient sampling, rather than in the coordination variables exchanged during distributed optimization. Therefore, there is a research gap in investigating noise effects on the learning accuracy. Furthermore, unlike classical neural networks, quantum circuits process data and learn nonlinear mappings in a fundamentally different way. The effects of noises on quantum circuits require further examination. While simple classification tasks have been studied in~\cite{watkins2023quantum}, complex nonlinear problems such as OPF remain unexplored.

In this paper, we address the research gaps above by proposing a privacy-preserving quantum neural network (QNN) for the probabilistic OPF problem. It leverages the power of quantum computing to accelerate the OPF calculation process. To prevent privacy leakage from the OPF solutions, we develop a ($\varepsilon,\delta$)-differentially private algorithm for QNN learning with proofs of privacy guarantees. Our main contributions are as follows:
\begin{itemize}
\item \textbf{A QNN-based method for probabilistic OPF problem}: We propose a QNN-based method for fast probabilistic OPF calculation. To enhance the expressive power, we introduce a strongly entangled variational layer within the variational quantum circuit. Compared to classical NNs, the QNN demonstrates superior approximation accuracy while reducing the number of parameters by 90\%.
\item \textbf{A differentially private QNN learning algorithm}: To train the proposed QNN  for solving OPF while preserving privacy, we develop a ($\varepsilon,\delta$)-differentially private algorithm with theoretical guarantees. By injecting controlled Gaussian noise into the quantum gradient update process, the proposed algorithm prevents adversaries from inferring the power usage patterns of customers. Moreover, the proposed QNN is noise-resilient, i.e., when adding noise at a large scale, the model updates steadily without issues such as gradient exploding in classical NNs.
\end{itemize}

\textit{Paper Organization:} Section~\ref{sec:preliminary} introduces the preliminaries of the probabilistic OPF problem and quantum computing. Section~\ref{sec:qnn} presents the proposed QNN following the discussion of the probabilistic OPF problem. To prevent sensitive information from being revealed by the QNN, Section~\ref{sec:dp} proposes the private learning algorithm with an $(\varepsilon,\delta)$-differential privacy guarantee. Case studies are conducted in Section~\ref{sec:case-study}, and the conclusion is provided in Section~\ref{sec:conclusion}.

\section{Preliminaries}
\label{sec:preliminary}
\subsection{Probabilistic Optimal Power Flow}
We consider a radial distribution grid with controllable distributed energy resources (DERs). A distribution system operator (DSO) is responsible for managing the DERs and controlling the power supply from the main grid, ensuring that grid security constraints are satisfied. The grid is represented as a graph $\mathcal{G}=(\mathcal{B},\mathcal{L})$, where $\mathcal{B}$ is the set of buses and $\mathcal{L}$ is the set of lines. Each bus in $\mathcal{B}$ is indexed by $i \in \{1, \dots, N\}$ and each line in $\mathcal{L}$ is denoted as $l = (j, k)$. Suppose each controllable DER sited at bus $i$ is with cost coefficient $c_i$, then the deterministic OPF problem is formulated as:
\begin{align}
\quad & \min_{\ell,P,Q,v} \sum_{i \in \mathcal{B}} c_{i} P^G_i \label{eq:obj-opf}\\
    \text{s.t.} \quad & P_{jk} - p_k - r_{jk} \ell_{jk} - \sum_{i:(k,i) \in \mathcal{L}} P_{ki} = 0, \quad \forall (j,k) \in \mathcal{L}, \label{eq:p-balance}\\
    & Q_{jk} + q_k - x_{jk} \ell_{jk} - \sum_{i:(k,i) \in \mathcal{L}} Q_{ki} = 0, \quad \forall (j,k) \in \mathcal{L}, \label{eq:q-balance}\\
    & v_{k} = v_{j} - 2(r_{jk} P_{jk} + x_{jk} Q_{jk}) \nonumber \\
    & \quad\quad\quad\quad + (r_{jk}^2 + x_{jk}^2) \ell_{jk}, \quad \forall (j,k) \in \mathcal{L}, \label{eq:voltage}\\
    & \ell_{jk} = \frac{P_{jk}^2 + Q_{jk}^2}{v_{j}}, \quad \forall (j,k) \in \mathcal{L}, \label{eq:current}\\
    & \underline{p}_i \leq p_{i} \leq \bar{p}_i, \quad \underline{q}_i \leq q_{i} \leq \bar{q}_i, \quad \forall i \in \mathcal{B}, \\
    & 0 \leq \ell_{jk} \leq \bar{\ell}_{jk}, \quad \forall (j,k) \in \mathcal{L}, \\
    & \underline{v}_i \leq v_{i} \leq \bar{v}_i, \quad \forall i \in \mathcal{B},
\end{align}
where the objective is to minimize the total operation cost subject to branch flow equations~\eqref{eq:p-balance}-\eqref{eq:current} adopted in~\cite{farivar2013branch}. $P^G_i$ denotes the generation power of controllable DER sited in bus $i$. The power $p_{i}$/$q_{i}$ represents the active/reactive power at bus $i$, $P_{jk}$/$Q_{jk}$ is the active/reactive power flow on line $l = (j,k) \in \mathcal{L}$, $r_{jk}$/$x_{jk}$ denotes the resistance/reactance of line $l = (j,k)$, $\ell_{jk}$ is the squared current magnitude of line $l = (j,k)$, $v_{j}$ is the squared voltage magnitude at bus $j \in \mathcal{B}$, and $\underline{\bullet}$ and $\overline{\bullet}$ represent the lower and upper bounds of the respective variables.

Suppose there are wind and solar generation in the distribution grid. To account for the uncertainties inherent in wind and solar generation, as well as load demand variability in the distribution grid, we extend the deterministic OPF into a probabilistic OPF formulation as follows:
\begin{align}
% \min_{\mathbf{u}}~ & \mathbb{E}_{\boldsymbol{\xi}} [ f(\mathbf{u}, \boldsymbol{\xi}) ], \label{eq:prob_opf}\\
% \mbox{s.t.}~ & h(\mathbf{u}, \boldsymbol{\xi}) = 0, \label{eq:prob_opf_const1}\\
% ~ & g(\mathbf{u}, \boldsymbol{\xi}) \leq 0, \label{eq:prob_opf_const2}
\mathbb{E}_{\boldsymbol{\xi}}& [\min_{\mathbf{u}}f(\mathbf{u}, \boldsymbol{\xi}) ], \label{eq:prob_opf}\\
\mbox{s.t.}~ & h(\mathbf{u}, \boldsymbol{\xi}) = 0, \label{eq:prob_opf_const1}\\
~ & g(\mathbf{u}, \boldsymbol{\xi}) \leq 0, \label{eq:prob_opf_const2}
\end{align}
where $\boldsymbol{\xi}$ represents the set of random variables capturing system uncertainties (i.e., solar generation, wind generation, and load demand), while $\boldsymbol{u}$ denotes the same set of decision variables as in the deterministic OPF. The objective function $\mathbb{E}_{\boldsymbol{\xi}} [\min_{\mathbf{u}}f(\mathbf{u}, \boldsymbol{\xi}) ]$ represents the expected minimum operational cost, where the operational cost is minimized for each uncertain scenario, and the expectation is taken over the distribution of these optimal operational costs. Constraint~\eqref{eq:prob_opf_const1} represents the branch flow equations, and constraint~\eqref{eq:prob_opf_const2} enforces the security limits on $\boldsymbol{u}$.

A classical method to solve the above probabilistic OPF problem is by the Monte Carlo simulation~\cite{mooney1997monte}. It draws samples of uncertain variables from the distributions, solves the deterministic OPF, and then aggregates the results. Despite the accuracy, Monte Carlo methods are computationally intensive. To accelerate the calculation of each sample, we leverage the computational power from quantum computing.

\subsection{Basic Concepts of Quantum Computing}
Quantum computing processes the information with \textit{qubit}, i.e., the quantum bit. Unlike classical bits, which can only exist in a distinct state of either 0 or 1, qubits can exist in both states simultaneously due to the property of quantum superposition.
%The state of a single qubit is shown in Fig.~\ref{fig:bloch_sphere}.
Mathematically, the state of a single qubit can be represented as:
\begin{equation}
|\psi\rangle = \alpha |0\rangle + \beta |1\rangle,
\end{equation}
where $|0\rangle = \begin{pmatrix}1 \\ 0\end{pmatrix}$ and $|1\rangle = \begin{pmatrix}0 \\ 1\end{pmatrix}$ are the basis states in Hilbert space and $\alpha$, $\beta \in \mathbb{C}$ are complex amplitudes satisfying the normalization condition $|\alpha|^2 + |\beta|^2 = 1$. 

\begin{figure*}
    \centering
    \includegraphics[width=\linewidth]{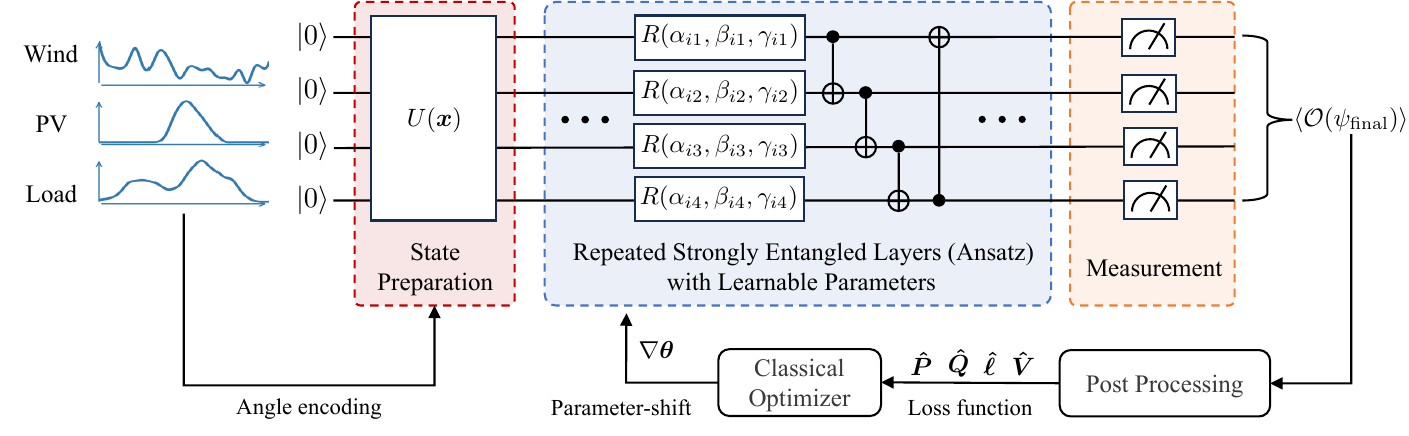}
    \caption{The proposed variational quantum circuit-based method for OPF solution learning.}
    \label{fig:method}
\end{figure*}

% \begin{figure}[H]
%     \centering
%     \includegraphics[width=0.6\linewidth]{figs/bloch.pdf}
%     \caption{Schematic representation of a qubit state on the Bloch's sphere. Each point on the sphere represents a quantum state $|\psi\rangle = \cos\frac{\theta}{2}|0\rangle + e^{i\varphi}\sin\frac{\theta}{2}|1\rangle$, which is in a superposition of states $|0\rangle$ and $|1\rangle$.}
%     \label{fig:bloch_sphere}
% \end{figure}
When solving practical problems, we use multiple qubits together by creating correlations between them. Such correlations are called \textit{entanglement}. When qubits are entangled with each other, operations on any one of the qubits perform corresponding parallel operations on the other qubits. Therefore, this entanglement contributes to the potential exponential speedup that quantum computers may possess over classical computers for certain algorithms. 

The quantum operations are performed by quantum gates. Quantum gates are unitary matrices in Hilbert space. For example, an elementary quantum gate, Pauli-X gate is denoted by $\begin{pmatrix}
0 & 1 \\ 1 & 0   
\end{pmatrix}$, which flips the state of a qubit. When applying quantum gates $U$ on the qubits, the quantum states unitarily change from initial state $|0\rangle$ to a new state $|\psi\rangle = U|0\rangle$, which is the unitary evolution of qubits. By designing appropriate quantum gates, qubits evolve toward the expected state, which can then be measured to collect the final result.

\section{Quantum Neural Network-Based OPF Approximation}
\label{sec:qnn}
To accelerate OPF problem-solving, we propose a quantum neural network-based method for OPF approximation.

% \subsection{Variational Quantum Circuits Learning}
Variational quantum circuit (VQC) is the ``neural network'' in quantum ML. It is a parameterized quantum circuit where multiple qubits evolve to perform a task. Unlike standard quantum circuits, VQCs include quantum gates with learnable parameters, which can be optimized iteratively based on a predefined loss function.

Consider the uncertainties in the distribution grid, such as wind generation $\boldsymbol{P}_{\textrm{wind}}$, solar generation $\boldsymbol{P}_{\textrm{solar}}$, and customer load demand $\boldsymbol{P}_{\textrm{load}}$ at various nodes. Given the input $\boldsymbol{x}= (\boldsymbol{P}_{\textrm{wind}}, \boldsymbol{P}_{\textrm{solar}}, \boldsymbol{P}_{\textrm{load}})$, the VQC aims to approximate the OPF solution $\boldsymbol{y}^* = (\boldsymbol{P}, \boldsymbol{Q}, \boldsymbol{\ell}, \boldsymbol{V})$. To learn the mapping $\boldsymbol{x} \rightarrow \boldsymbol{y}^*$, classical NNs use a feed-forward approach: for each sample $\boldsymbol{x}$, the NN predicts $\boldsymbol{\hat{y}}$, calculates the loss $\mathcal{L}$ between $\boldsymbol{\hat{y}}$ and the true output $\boldsymbol{y}^*$, and updates the parameters using the gradients $\nabla \theta_{\textrm{NN}}$ from $\mathcal{L}$. Similarly, the VQC follows the same paradigm but requires additional conversions between quantum and classical information.

We illustrate the proposed QNN-based OPF learning method in Figure~\ref{fig:method}. The first step is to encode the classical input $\boldsymbol{x}$ into the quantum system, i.e., the state preparation. Here, we use the angle encoding to encode the input values into the relative phase in the quantum states:
\begin{align}
|U_{\textrm{enc}}\rangle = U(\boldsymbol{x})|0\rangle^{\otimes n} =\bigotimes_{i=1}^n \left(\cos\left(\frac{x_i}{2}\right)|0\rangle +\sin\left(\frac{x_i}{2}\right)|1\rangle\right).
\end{align}

After encoding, repeated variational layers (also called ansatz) evolve the prepared quantum system $|U_{\textrm{enc}}\rangle$ into the target system $|\psi_{\textrm{final}}\rangle = U(\boldsymbol{\theta})|U_{\textrm{enc}}\rangle$. This process is analogous to the hidden layers in classical NNs, where the uncertain variables in $\boldsymbol{x}$ are mapped to the OPF solution $\boldsymbol{y}^*$. The quantum gates include parameterized rotation gates and controlled-NOT (CNOT) gates: the learnable parameters $\boldsymbol{\theta}$ control the evolution and are optimized during the learning process; the controlled gates entangle qubits, enabling nonlinear transformations and thus increasing the expressive power. Moreover, the number of layers can also be tuned: the more layers, the larger the expressive power. To further improve this capability, we incorporate a strongly entangled layer~\cite{schuld2020circuit}. This layer consists of qubit rotations followed by adjacent pairwise CNOT gates, thereby generating states that strongly entangle each qubit and provide significant non-linear expressiveness.

To obtain the predicted OPF solution $\boldsymbol{\hat{y}}$ from the final quantum system $|\psi_{\textrm{final}}\rangle$, we need to convert the quantum information into the classical information. This is achieved by the measurement operation. Before the measurement, the quantum system $|\psi_{\textrm{final}}\rangle$ is in the superposition of multiple states, with each state representing one possible output of OPF solution $\boldsymbol{\hat{y}}$. For example, a quantum state $|\psi \rangle = \sum_{i=0}^{2^n-1}\sqrt{p_i}|i\rangle$ with $n$ qubits and $||p||_2 = 1$ represents $n$ components of a OPF solution with $2^n$ possible combinations. However, during the general measurement process, each qubit can only collapse onto 0 or 1, which cannot represent the continuous value of an OPF solution, e.g., the power flow or the voltage magnitude. Therefore, here we calculate the expectation of eigenvalues in each qubit, i.e., the expectation value of each component in OPF solution $\boldsymbol{\hat{y}}$. Mathematically, this process can be expressed by
\begin{align}
\boldsymbol{\hat{y}} = \langle \psi_{\textrm{final}} | \mathcal{O} | \psi_{\textrm{final}} \rangle = \langle \mathcal{O}(\psi_{\textrm{final}}) \rangle,
\end{align}
where $\mathcal{O}$ is the observable operator (i.e., a Hermitian matrix with real eigenvalues).

After measurement, we obtain the predicted OPF solution $\boldsymbol{\hat{y}}$. By evaluating the loss between the predicted value $\boldsymbol{\hat{y}}$ and true value $\boldsymbol{y}^*$, we calculate the gradient $\nabla \boldsymbol{\theta}$ of learnable parameters $\boldsymbol{\theta}$. Then, we update the parameters with a classical optimizer iteratively to approximate the true OPF solution $\boldsymbol{y}^*$.

\section{Differentially Private Learning for Optimal Power Flow}
\label{sec:dp}

% privacy issue in POPF
In the previous section, the VQC learns from the OPF solution $\boldsymbol{y}^*$ from~\eqref{eq:obj-opf} and produces a solution $\boldsymbol{\hat{y}}$ that is as accurate as possible. However, if $\boldsymbol{\hat{y}}$ is exposed, an attacker could potentially decode power consumption patterns of consumers using methods like non-intrusive load monitoring (NILM)~\cite{wang2020privacy}. To be specific, voltage and power flow measurements in $\boldsymbol{\hat{y}}$, can potentially reveal residential customer activities (such as electric vehicle charging) and industrial customer production schedules, creating significant privacy risks.

To address this privacy concern, in this section, we develop a privacy-preserving learning algorithm for VQC. The algorithm is proved to be ($\varepsilon,\delta$)-differential privacy for given load datasets.

\subsection{Differential Privacy}
To quantify and control the privacy risks of the customer loads, we adopt the framework of \textit{differential privacy}~\cite{dwork2014algorithmic}. It protects privacy by ensuring the outputs from two similar inputs are in-distinguishable. Before introducing the mechanism of differential privacy, we briefly introduce the meaning of ``similarity''. 

Suppose a load dataset is represented by $D \in \mathbb{R}^n$ with $i$-th element $d_i$ describing the active load of node $i$. For two load datasets $D$ and $D'$ that differ only in one element, they hold the following \textit{adjacency} relation $D \sim_\mu D'$:
\begin{align}
     \exists i, \mbox{s.t.}~ |d_i- d_i'| \le \mu ~\mbox{and}~ d_j=d_j',\forall j \ne i,
\end{align}
% mechanism of differential privacy
where $\mu$ is the scalar measuring the distance between two adjacent datasets. 

For a non-private OPF algorithm, the corresponding outputs given two adjacent load datasets, $D$ and $D'$, are distinguishable, leaving exploitable information for an adversary to reverse-engineer the original load dataset. Differential privacy defines the degree of privacy leakage in a probabilistic sense.
\begin{definition}[$(\varepsilon, \delta)$-differential privacy~\cite{dwork2014algorithmic}]
A randomized OPF algorithm $\mathcal{M}: \mathcal{D} \rightarrow \mathcal{S}$ from the domain $\mathcal{D}$ to domain $\mathcal{S}$ is $(\varepsilon,\delta)$-differentially private if, for all $s \subseteq \mathcal{S}$ and for any adjacent pair $D \sim_\mu D' \in \mathbb{R}^n$, the following condition holds:
\begin{align}
    \mathbb{P}[\mathcal{M}(D) \in s] \le \text{exp}(\varepsilon) \mathbb{P}[\mathcal{M}(D') \in s] + \delta.
\end{align}
\end{definition}
\noindent Here, the domain $\mathcal{D}$ in OPF problem~\eqref{eq:obj-opf} includes feasible load datasets, solar and wind generation, and the domain $\mathcal{S}$ denotes the feasible region. The level of privacy protection is defined by parameters ($\varepsilon,\delta$). $\varepsilon$ controls the maximum distance between the outputs of two adjacent datasets, and $\delta$ denotes the probability of privacy leakage in the worst case. Therefore, the smaller ($\varepsilon,\delta$) provides stronger privacy protection.

\subsection{Differentially Private Learning for Variational Quantum Circuit}

To integrate differential privacy into the learning process of VQC, we add Gaussian noise into gradients $\nabla_\theta \mathcal{L}(\boldsymbol{\theta},\boldsymbol{x}_i)$ of parameterized quantum gates and clip the gradient of each sample by a pre-defined $\ell_2$ norm $C$.

Algorithm~\ref{alg:dp_qnn} outlines the differentially private VQC learning process. For each OPF sample, it encodes the input data into a quantum state, evolves it through the VQC, and measures the final quantum system to obtain the predicted OPF solution. After that, a gradient is calculated based on the prediction loss and then is clipped based on a norm $C$. By clipping, the influence of each individual sample on the gradient is bounded to protect privacy. Then, the Gaussian noises controlled by scale $\sigma$ are added to the clipped gradients before updating the VQC parameters $\boldsymbol{\theta}$. This process is repeated for multiple epochs, leading to a privacy-preserving VQC that can approximate OPF solutions while protecting individual load data.

\begin{algorithm}[H]
\caption{Differentially Private VQC Learning for OPF}
\label{alg:dp_qnn}
\begin{algorithmic}[1]
\REQUIRE OPF dataset $\{ (\boldsymbol{x}_i, \boldsymbol{y}^*_i): i=1,\dots,N\}$, loss function $\mathcal{L(\boldsymbol{\theta})} = \mathcal{L}(\boldsymbol{\theta}, \boldsymbol{\hat{y}}_i, \boldsymbol{y}^*_i)$, learning rate $\eta$, batch size $B$, clipping norm $C$, Gaussian noise multiplier $\sigma$, initial VQC with parameters $\boldsymbol{\theta}_0$, observable operator $\mathcal{O}$, total epochs $T$
\WHILE{epoch index $e \leq$ total epochs $T$}
\FOR{each batch $\mathcal{B} \subset \mathcal{D}$}
    \FOR{each OPF sample $(\boldsymbol{x}_i, \boldsymbol{y}_i) \in \mathcal{B}$}
        \STATE Prepare quantum system $|\psi_{\textrm{enc}}\rangle$ by encoding OPF data $\boldsymbol{x}_i$
        \STATE Evolve $|\psi_{\textrm{enc}}\rangle$ to the final quantum system $|\psi_{\textrm{final}}\rangle$ through repeated variational layers with parameters $\boldsymbol{\theta}$
        \STATE Calculate the predicted OPF solution $\boldsymbol{\hat{y}}_i$ by measurement operation $\boldsymbol{\hat{y}}_i = \langle|\psi_{\textrm{final}}|\mathcal{O}|\psi_{\textrm{final}}\rangle$ with observable operator $\mathcal{O}$
        \STATE Compute gradient $\boldsymbol{g}_i = \nabla_{\boldsymbol{\theta}} \mathcal{L}(\boldsymbol{\theta}, \boldsymbol{\hat{y}}_i, \boldsymbol{y}_i)$
        \STATE Clip gradient: $\boldsymbol{g}_i \leftarrow \boldsymbol{g}_i \cdot \min \left( 1, \frac{C}{\| \boldsymbol{g}_i \|_2} \right)$
    \ENDFOR
    \STATE Add noise: $\tilde{\boldsymbol{g}} = \frac{1}{B}  (\sum_{i=1}^{B}\boldsymbol{g}_i + \mathcal{N}(0, \sigma^2 C^2 \boldsymbol{I}))$
    \STATE Update parameters: $\boldsymbol{\theta} \leftarrow \boldsymbol{\theta} - \eta \tilde{\boldsymbol{g}}$
\ENDFOR
\ENDWHILE
\end{algorithmic}
\end{algorithm}
\subsection{Proof of Privacy Guarantees}

The privacy guarantee of Algorithm~\ref{alg:dp_qnn} is described as the following theorem.

\begin{theorem}[Privacy Guarantees]
For any adjacent pairs of load demand datasets $(D, D')$, if $\varepsilon, \delta \in (0,1)$, $\sigma \geq \frac{\sqrt{2\log(1.25)/\delta}}{\epsilon}$, and $\delta' > 0$, then, Algorithm~\ref{alg:dp_qnn} is $(\varepsilon', \frac{TB}{N}\delta +\delta')$-differentially private, where
\begin{align}
\varepsilon' = \sqrt{2T\ln (1/\delta')}\cdot  \frac{B\varepsilon}{N} + \frac{TB\varepsilon}{N}(e^{B\varepsilon/N} -1).
\end{align}
\end{theorem}

\textit{Proof:} In each step of Algorithm~\ref{alg:dp_qnn}, the noise following $\mathcal{N}(0, \sigma^2C^2\boldsymbol{I})$ are added on the average gradient of each batch. For each batch, we compute the per-sample gradients and clip them to differ at most $C$, which bounds the sensitivity of gradient $\Delta$ within $C$. By Gaussian Mechanism~\cite{dwork2014algorithmic}, if we choose $\sigma \geq \frac{\sqrt{2\log(1.25)/\delta}}{\epsilon}$, the algorithm is ($\varepsilon,\delta$)-differentially private. Since in each step, we sample one batch of size $B$ from $N$ data samples with the sampling rate $\frac{B}{N}$. From the theorem of privacy amplification by uniform subsampling~\cite{balle2018privacy}, the resulting algorithm achieves ($\frac{B}{N}\varepsilon, \frac{B}{N}\delta$)-differentially privacy. Since there are $T$ epochs, we apply the privacy-preserving algorithm for $T$ times. From the advanced composition theorem~\cite{dwork2014algorithmic}, i.e., if the algorithm is composed of $T$ algorithms with each providing ($\varepsilon, \delta)$-differential privacy, the composed algorithm is $(\varepsilon'', T\delta + \delta'')$-differential private for any $\delta'' >0$ with $\varepsilon'' = \sqrt{2T\ln(1/\delta'')}\cdot \varepsilon + T\varepsilon(e^\varepsilon - 1)$. 

Hence, the Algorithm~\ref{alg:dp_qnn}, which composes $T$ steps with each providing ($\frac{B}{N}\varepsilon, \frac{B}{N}\delta$)-differentially privacy, achieves $(\varepsilon', \frac{TB}{N}\delta +\delta')$-differentially privacy. This completes the proof.$\hfill \square$

\section{Case Study}
\label{sec:case-study}
\begin{figure}[H]
    \centering
    \includegraphics[width=0.9\linewidth]{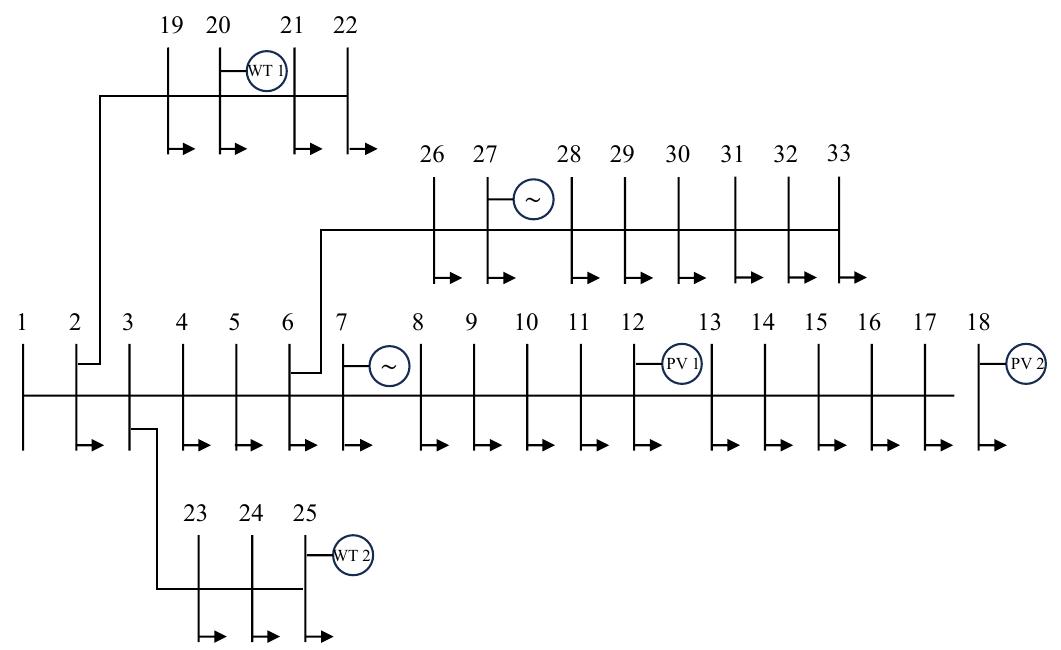}
    \caption{Topology of modified IEEE 33-bus power distribution system.}
    \label{fig:grid}
\end{figure}

Numerical experiments are conducted on the modified IEEE 33-bus power distribution system. As illustrated in Figure~\ref{fig:grid}, the system integrates 2 distributed generators (DGs), 2 Photovoltaic (PVs), and 2 wind turbines (WTs). The active and reactive power limits are set as follows: $\underline{p}_i=0$ MW, $\bar{p}_i=8$ MW, $\underline{q}_i=0$ MVar, and $\bar{q}_i=2$ MVar. The relaxed AC-OPF problem~\eqref{eq:obj-opf} is coded in MATLAB using the YALMIP interface and solved with GUROBI software. The QNN is implemented using the Pennylane library and JAX library. All experiments are conducted on a MacBook with the M3 Max Apple Silicon chip. The test hyperparameters are presented in Table~\ref{tab:hyperparameters}.

\begin{figure*}
    \centering
    \includegraphics[width=\linewidth]{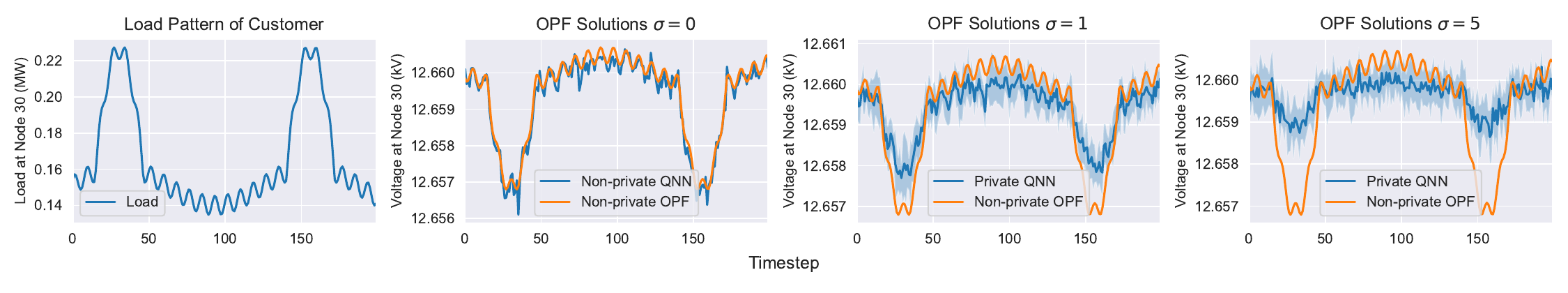}
    \caption{Voltage magnitude comparisons between differentially private QNN and non-private OPF under load changes. The solid line represents the mean predicted value from 10 measurements, each consisting of 100 shots. The shaded area denotes the mean value plus or minus the standard deviation.}
    \label{fig:example}
\end{figure*}

The input features of the QNN include uncertain variables of 2 WTs, 2 PVs, and load demand from 1 customer. The output is the voltage magnitude at node $i$ due to the qubit limitation in the Noisy Intermediate-Scale Quantum (NISQ) era~\cite{preskill2018quantum}.

\begin{table}[H]
\centering
\caption{Test Hyperparameters}
\label{tab:hyperparameters}
\resizebox{\columnwidth}{!}{
\begin{tabular}{cccc}
\hline
Hyperparameters        & Value & Hyperparameters              & Value        \\ \hline
learning rate ($\eta$) & 0.05 & clipping norm ($C$)          & 1            \\
batch size             & 32    & number of variational layers &  10           \\
qubits                 & 5     & optimizer & Adam \\ \hline
\end{tabular}%
}
\end{table}

\subsection{Privacy Guarantees}
We first provide an illustrative example to demonstrate how an adversary detects the load pattern if the voltage measurements are exposed. Assume a customer at node 30 has an atypical load pattern representing the production technology. Its pattern under different timestep $t$ is formulated with the following three periodic components~\cite{dvorkin2020differentially}:
\begin{align*}
   p(t) = \max \left\{ \sin \frac{5}{10^2}t, \frac{7}{10} \right\} + \frac{5}{10^2} \sin \frac{5}{10^2}t + \frac{25}{10^3} \sin \frac{75}{10^2}t.
\end{align*}

The non-private OPF solution leaks the sensitive load information through the voltage readings, as displayed in Figure~\ref{fig:example}.
The left figure presents the load pattern of the customer and the remaining figures show the voltage magnitudes at the same node. From left to right, in the second figure, we can see that the non-private OPF and non-private QNN demonstrate an informative ``reverse'' pattern of the load demand, disclosing the power consumption pattern of the customer. When the Gaussian noises are added to the training process, i.e., $\sigma \neq 0$, the solution from private QNN deviates from the non-private one and fluctuates more mildly to protect the load pattern. When noise with a larger scale $\sigma$ is introduced, the fluctuation is smaller, leading to stronger privacy protection.

\subsection{Probabilistic OPF Solutions}

To assess the accuracy of the probabilistic OPF solution generated by the private QNN, we computed the mean and standard deviation based on 1,000 samples. These samples were obtained by sampling the WT generation from a Weibull distribution $\textrm{Weibull}(1, 4.8)$, PV generation from a Beta distribution $\textrm{Beta}(2, 5)$, and load demand from a normal distribution $\mathcal{N}(0, 0.3)$. These samples were then solved deterministically to collect the aggregate statistics.

Table \ref{tab:v30-POPF} presents a comparison of the voltage magnitude at Node 30 across various perturbation levels ($\sigma = 0$, $1$, $5$, and $10$). The reference values, i.e., the mean value of 12.6592 kV and standard deviation value of 0.0009 kV, were obtained via the Monte Carlo method. In the absence of noise ($\sigma = 0$), the calculated statistics closely align with the reference values. As the noise scale increases, representing a greater level of privacy protection, the statistical results deviate more significantly from the reference mean. Specifically, the error percentage for the mean voltage remains within 0.0019\%, indicating that the introduction of noise by the private QNN causes only minimal alterations to the voltage magnitudes, thereby allowing for accurate estimation of aggregate statistics. However, the standard deviation exhibits larger deviations as the noise restricts the model’s sensitivity to minor variations. Nonetheless, by selecting an appropriate noise scale, e.g. from $\sigma = 0$ to $\sigma = 1$, a balanced trade-off can be achieved between the accuracy in estimating the uncertainties and the privacy protection of the sensitive load patterns.

\begin{table}[H]
\centering
\caption{POPF Accuracy of Private QNN}
\label{tab:v30-POPF}
\resizebox{\columnwidth}{!}{%
\begin{tabular}{lccccc}
\hline
\multicolumn{6}{c}{Voltage Magnitude (kV) Results at Node 30 ($N = 1000$)}                                \\ \hline
\multirow{2}{*}{Quantities} &
  \multirow{2}{*}{Reference} &
  \multirow{2}{*}{$\sigma = 0$} &
  \multirow{2}{*}{$\sigma = 1$} &
  \multirow{2}{*}{$\sigma = 5$} &
  \multirow{2}{*}{$\sigma = 10$} \\
                                           &                  &          &          &          &          \\ \hline
Mean                                       & 12.6592          & 12.6592  & 12.6592  & 12.6592  & 12.6594  \\
$\varepsilon^{V_{30}}_\text{mean}(\%)$     & \textbackslash{} & $\leq 10^{-5}$   & 0.0002   & 0.0008   & 0.0019   \\ \hline
STD                                        & 0.0009           & 0.0009   & 0.0007   & 0.0006   & 0.0005   \\
$\varepsilon^{V_{30}}_\text{STD}(\%)$      & \textbackslash{} & 1.3087   & 18.8674  & 36.6288  & 37.9414  \\ \hline
% Skewness                                   & 0.0973           & -0.1859  & -0.2739  & -0.1799  & -0.3131  \\ \hline
% $\varepsilon^{V_{30}}_\text{skewness}(\%)$ & \textbackslash{} & 291.0575 & 381.5006 & 284.8867 & 421.7816 \\ \hline
% Kurtosis                                   & 0.1133           & -0.2167  & 0.1462   & -0.0303  & -0.1502  \\
% $\varepsilon^{V_{89}}_\text{kurtosis}(\%)$ & \textbackslash{} & 291.2642 & 29.0452  & 126.7451 & 232.5688 \\ \hline
\end{tabular}%
}
\end{table}

\subsection{Comparison with Classical ML}

To demonstrate the advantage of quantum ML over classical ML, we implement a multilayer perceptron (MLP) and compare the accuracy between the proposed QNN and a classical MLP network. The MLP is developed using the PyTorch library and consists of two hidden layers, each containing 32 units. Both models are trained for 1,000 epochs. When noise is introduced, both models utilize Algorithm~\ref{alg:dp_qnn} to preserve the privacy of load demands. The evaluation metric is the coefficient of determination, i.e., $R^2$, which quantifies the degree to which a model's predictions align with observed data, with values closer to 1 indicating a better fit.

Due to the limited availability of real quantum hardware, we adopt the formula proposed by Guerreschi et al.~\cite{guerreschi2019qaoa} to calculate the quantum computational time, including the execution time and the measurement time:
\begin{align*}
T = T_p + T_G \times D + T_M,
\end{align*}
where $T_p$ denotes the time for state preparation, $T_G$ represents the average duration of a single quantum gate, $D$ is the maximal circuit depth (number of sequential quantum gates), and $T_M$ is the measurement time. The time scales are set as $T_p + T_M = 1\ \mathrm{\mu}\mathrm{s}$ and $T_G = 10\ \mathrm{ns}$, following the parameters in literature~\cite{zhou2024carbon}.

\begin{table}[H]
\centering
\caption{OPF Accuracy Comparison between MLP and QNN}
\label{tab:MLP-QNN}
\resizebox{\columnwidth}{!}{%
\begin{tabular}{ccccccc}
\hline
\multicolumn{7}{c}{Voltage Magnitude (kV) Results at Node 30 ($N = 1000$)}                                \\ 
\hline
Model & Calculation Time        & \#Parameters & $\sigma = 0$ & $\sigma = 1$ & $\sigma = 2$ & $\sigma = 5$ \\ \hline
MLP   & $1.12 \times 10^{-4}$ & 1281         & 0.971        & -0.065       & -0.095       & -22291.787   \\
QNN   & $\mathbf{1.76 \times 10^{-6}}$ & \textbf{165}          & \textbf{0.977}        & \textbf{0.739}        & \textbf{0.291}        & \textbf{0.113}        \\ \hline
\end{tabular}%
}
\end{table}

Table~\ref{tab:MLP-QNN} presents a comparison of the accuracy in predicting the voltage magnitude at Node 30 across 1,000 samples. Under various noise scales, the QNN consistently outperforms the MLP. In the absence of noise, the obtained $R^2$ scores are comparable, with the QNN achieving 0.977 and the MLP achieving 0.971. However, when Gaussian noise is incorporated into the training process, the QNN demonstrates a clear advantage over the MLP. Specifically, at noise levels of $\sigma = 1$ and $\sigma = 2$, the MLP yields negative $R^2$ scores, whereas the QNN maintains positive $R^2$ values, indicating the MLP's inability to effectively forecast the voltage magnitude under noisy conditions. This degradation is attributed to the accumulated noises, which hinder the MLP's forecasting capability. Notably, at a noise scale of $\sigma = 5$, the MLP exhibits a large negative $R^2$ value of -22,291.787, while the QNN still maintains a positive $R^2$ of 0.113. Furthermore, the QNN has shorter computation times and utilizes significantly fewer parameters, i.e., 165 versus 1,281, as shown in the second and third columns of Table~\ref{tab:MLP-QNN}, further highlighting the advantages of quantum ML over classical approaches.

\begin{figure}[H]
    \centering
    \includegraphics[width=0.8\linewidth]{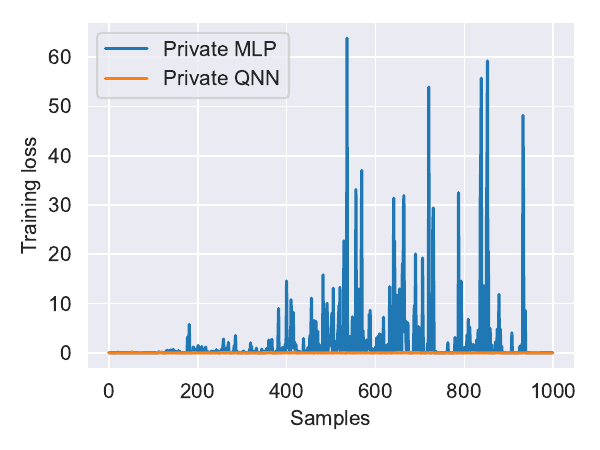}
    \caption{Training loss of QNN and MLP when $\sigma=1$. The classical MLP suffers from training stability while QNN is DP noise-resilient.}
    \label{fig:loss}
\end{figure}

To investigate the underlying reasons for the distinct performance between the QNN and MLP, we analyze the loss values during the training process. Figure~\ref{fig:loss} illustrates the case when $\sigma = 1$. The graph reveals that the training loss of the MLP is highly unstable; although most loss values are around zero, some instances exhibit very large loss values. A potential reason is that the accumulated noises on gradients dominate the learning process and thus destroy the subsequent learning performance. In contrast, the QNN maintains stable training and is resilient to noise, highlighting the advantage of the QNN when incorporating differential privacy into the training process.

\section{Conclusion}
\label{sec:conclusion}
Volatile and stochastic renewables and loads call for efficient and accurate calculation of probabilistic optimal power flow (OPF). Quantum computing presents promising speedup for approximating OPF efficiently by leveraging the quantum superposition and entanglement. However, the exposure of non-private OPF can potentially reveal the sensitive load demand pattern, raising privacy concerns. To this end, we propose a differentially private quantum neural network (QNN)-based method for the probabilistic OPF problem. By introducing carefully calibrated Gaussian noises into the training process, it is proven to achieve ($\varepsilon,\delta$)-differential privacy, leading to strong privacy protection. Moreover, a strongly entangled layer is employed to further improve the expressive power of variational quantum circuits. The case study validates that the proposed method preserves private load information while still maintaining an accurate estimation of the statistics in probabilistic OPF. Moreover, compared to the private classical neural network, the proposed private QNN achieves significantly higher accuracy and is more noise-resilient.

\section{Acknowledgments}
This work is supported by the CUHK Strategic Partnership Award for Research Collaboration under Grant No. 4750467, the CUHK Direct Grant for Research under Grant No. 4055228, and the CUHK startup fund.

\nocite{*}
\bibliographystyle{IEEEtran}
\bibliography{ref}
\end{multicols*}

\end{document}